\documentclass{emulateapj}

\usepackage{amsmath}

\slugcomment{Draft version Oct 13, 2015}

\shorttitle{Raman Scattering Wings in AGNs}
\shortauthors{Chang et al.}

\begin{document}


\title{Formation of Raman Scattering Wings around H$\alpha$, H$\beta$ and Pa$\alpha$ 
in Active Galactic Nuclei}


\author{Seok-Jun Chang\altaffilmark{1},  
 Jeong-Eun Heo\altaffilmark{1}, 
Francesco Di Mille\altaffilmark{2}, Rodolfo Angeloni\altaffilmark{3}, 
Tali Palma\altaffilmark{4,5},
and Hee-Won Lee\altaffilmark{1}}
\altaffiltext{1}{Department of Physics and Astronomy, Sejong University, Korea}
\altaffiltext{2}{Las Campanas Observatory, Chile}
\altaffiltext{3}{AURA-GEMINI Observatory, Chile}
\altaffiltext{4}{Millennium Institute of Astrophysics, Chile}
\altaffiltext{5}{Pontificia Universidad Catolica de Chile, Chile}





\begin{abstract}
Powered by a supermassive black hole with an accretion disk, 
the spectra of active galactic nuclei (AGNs) are characterized by prominent emission lines including Balmer lines. The unification schemes of AGNs require the existence of a thick molecular torus that may hide the broad emission line region from the view of observers near the equatorial direction. In this configuration, one may expect that the far UV radiation from the central engine can be Raman scattered by neutral hydrogen to reappear around Balmer and Paschen emission lines which can be identified with broad wings. We produce H$\alpha$, H$\beta$ and Pa$\alpha$ wings using a Monte Carlo technique to investigate their properties. The neutral scattering region is assumed to be a cylindrical torus specified by the inner and outer radii and the height. While the covering factor of the scattering region affects the overall strengths of the wings, 
the wing widths are primarily dependent on the neutral hydrogen column density $N_{\rm HI}$ 
being roughly proportional to $N_{\rm HI}^{1/2}$. In particular, with $N_{\rm HI}=10^{23}{\rm\ cm^{-2}}$ the H$\alpha$ wings typically show a width $\sim 2\times 10^4{\rm\ km\ s^{-1}}$. We also find that H$\alpha$ 
and Pa$\alpha$ wing profiles are asymmetric with the red part stronger than the blue part and an opposite behavior is seen for H$\beta$ wings.
\end{abstract}





\section{Introduction\label{sec:intro}}

Active galactic nuclei (AGNs) are known to be powered by a supermassive 
black hole with an accretion disk. The spectra of AGNs are characterized 
by a nonthermal featureless continuum with prominent emission 
lines exhibiting a large range of ionization and excitation \citep[e.g.,][]{pet}. 
One classification of AGNs can 
be made based on the width of emission lines. Type 1 AGNs show broad permitted 
lines and semi-forbidden lines with a typical width of $5,000{\rm\ km\ s^{-1}}$. 
In addition to these broad emission lines they also show narrow forbidden lines 
with a width $\sim 500{\rm\ km\ s^{-1}}$. In contrast to this, type 2 AGNs 
exhibit only narrow emission lines encompassing both permitted and forbidden lines. 

\cite{blandford82} proposed that the location of the broad line region can be 
observationally constrained by monitoring
the flux of emission lines that vary in response to the changes of continuum flux.
The time delay is directly translated into the size of the broad emission line
region, which is also used to estimate the mass of the black hole \citep[e.g.,][]{bentz09, park2012}.
The reverberation mapping shows that the broad emission line region is located 
within $\sim 0.1{\rm\ pc}$ from the central engine, whereas the narrow emission 
lines do not exhibit flux variations correlated with the neighboring continuum
\citep[e.g.,][]{peterson93, diet12}. 
A typical AGN unification model proposes that an optically and geometrically thick 
component resides between the broad emission line region and the narrow 
emission line region, hindering the observers in the equatorial direction 
from viewing the broad line region. 

 Spectropolarimetry can be an efficient tool to verify this unification scheme, 
because we expect the broad emission lines may be seen in the polarized spectra 
in the presence of a scattering medium in the polar direction 
\citep[e.g.,][]{ant1993}. 
The prototypical type 2 Seyfert galaxy, NGC 1068, shows broad Balmer lines 
in the polarized flux \citep[e.g.,][]{ant1985}. A similar observation was made 
for the narrow line radio galaxy Cyg A by \cite{ogle97}, in which an extremely broad H$\alpha$ line 
was found in the polarized flux. 

Broad absorption line quasars (BALQs), constituting about 10 percent of quasars,
exhibit broad absorption troughs in the blue part 
 of permitted broad lines.
The exact nature of the absorbing media
being controversial, one suggestion is that broad absorption lines are formed
in the equatorial outflow that is driven radiatively by quasar luminosity
 \citep[e.g.,][]{mur1995}. In this case the broad troughs will not be completely
 black but filled partially by photons resonantly scattered in other lines of sight
\citep[e.q.,][]{lee1997}. 
Spectropolarimetry is also applied to
find polarized residual fluxes in the broad absorption troughs
in a number of BALQs \citep[e.g.,][]{cohen95}.

In the presence of an optically thick component, it is expected that far UV radiation
can be inelastically scattered by neutral hydrogen, which may result in 
scattered radiation in the visible and IR regions. Raman scattering by atomic hydrogen 
was first introduced by \cite{schmid89}, when he identified the broad features around
$6825 \,{\rm \AA}$ and $7082 \,{\rm \AA}$.  These mysterious broad features are found 
in about half of symbiotic stars, wide binary 
systems consisting of an active white dwarf
and a mass losing giant \citep[e.q.,][]{kenyon1986}. 
The broad 6825 and 7082 emission features are formed through Raman
scattering of O~VI$\lambda\lambda$1032 and 1038,
when the scattering hydrogen atom in the ground state before incidence
finally de-excites to the $2s$ state.

The cross sections of
Raman scattering for O~VI$\lambda\lambda$1032 and 1038 are
$6.6 \sigma_{Th}$ and $2.0 \sigma_{Th}$, respectively, where
$\sigma_{Th}=0.665\times10^{-24}{\rm\ cm^{2}}$ is the Thomson
scattering cross section.
Due to the small scattering cross section, the operation of Raman scattering by atomic
hydrogen requires a large amount of neutral hydrogen that is illuminated 
by the far UV emission source. However, the cross section increases
sharply as the incident wavelength approaches those of Lyman series transition 
of hydrogen due to resonance.

The energy conservation requires the wavelength $\lambda_o$ of the Raman
scattered radiation to be related to the incident wavelength $\lambda_i$ by
\begin{equation}
\lambda_o^{-1}=\lambda_i^{-1}-\lambda_\alpha^{-1},
\label{wavelength_rel}
\end{equation}
where $\lambda_\alpha$ is the wavelength of Ly$\alpha$. This relation
immediately leads to the following 
\begin{equation}
{\Delta\lambda_o\over\lambda_o}={\lambda_o\over\lambda_i}{\Delta\lambda_i\over\lambda_i},
\label{wavelength_rel1}
\end{equation}
which dictates that the Raman features have a large width broadened by the factor
$\lambda_o/\lambda_i$.
In the case of Raman scattering of O~VI$\lambda$1032, $\lambda_o/\lambda_i\simeq 6.6$
which explains the abnormally broad width exhibited in the Raman O~VI$\lambda$6825 feature.

In a similar way, Raman
scattering of continuum photons in the vicinity of Ly$\beta$ may form
broad features around H$\alpha$. This mechanism has been invoked to explain
broad H$\alpha$ wings prevalent in symbiotic stars \citep{lee2000,yoo2002}. 
Lee \& Yun (1998)
also discussed polarized H$\alpha$ through Raman scattering in active galactic 
nuclei, where neutral regions may be thicker than those found in symbiotic stars. 

The next section provides summary of the atomic physics involving Raman scattering
by atomic hydrogen. Subsequently we provide our model for computation of broad
Balmer and Paschen wings with our simulated results. A brief discussion is provided 
before conclusion.

\section{Atomic Physics of Raman Scattering by H~I\label{sec:atomic}}

The exact nature of the thick absorbing component surrounding
the broad emission line region in AGNs is still controversial. X-ray observation
can be an excellent tool to probe the physical properties of the intrinsic absorber
in AGNs. One such study performed using {\it Suzaku} by \cite{chiang2013} reported the column
density of hydrogen $N_{\rm H I}\sim 5\times 10^{23}{\rm\ cm^{-2}}$ 
in the type 2 quasar IRAS~09104+4109. In the presence of this thick neutral hydrogen,
the Rayleigh and Raman scattering optical depth for radiation in the vicinity 
Lyman series can be quite significant.

The scattering of light by an atomic electron is described by the second order
time dependent perturbation theory.
We consider an incident photon with angular frequency $\omega$ with the polarization
vector ${\boldsymbol\epsilon}^{\alpha}$ scattered by an electron in the initial state $A$,
which subsequently de-excites to the final state $B$ accompanied by the emission of 
an outgoing photon 
with angular frequency $\omega'$ with polarization
vector ${\boldsymbol\epsilon}^{\alpha'}$.
The energy conservation requires that the difference of the photon energy
\begin{equation}
\hbar(\omega-\omega')=E_B-E_A,
\end{equation}
where $E_A$ and $E_B$ are the energy of the
initial and final states $A$ and $B$, respectively. 

Depending on the sign of $E_B-E_A$, the emergent Raman lines are
classified into Stokes and anti-Stokes lines.
When $E_B-E_A>0$, we have a Stokes line, which
is less energetic than the incident radiation. If $E_B-E_A<0$, 
then the transition corresponds to an anti-Stokes line. 
In this work, all the transitions correspond to Stokes lines.

In this work, no consideration is made of the polarization of Raman wings, which is deferred to a future work.

The cross section for this interaction is given 
by the Kramers-Heisenberg formula
\begin{eqnarray}
{d\sigma\over d\Omega}
&=& r_0^2 \left( {\omega'\over\omega} \right)
\left| \delta_{AB}{{\boldsymbol\epsilon}^{\alpha}}\cdot{{\boldsymbol\epsilon}^{\alpha'}}
\right.
\nonumber \\
&-&{1\over m_e\hbar}
\sum_I \left({{({\bf p}\cdot {\boldsymbol\epsilon}^{(\alpha')})_{BI}
({\bf p}\cdot {\boldsymbol\epsilon}^{(\alpha)})_{IA}}\over{\omega_{IA}-\omega}} 
\right.
\nonumber \\ 
&-& \left.\left.  {{({\bf p}\cdot {\boldsymbol\epsilon}^{(\alpha)})_{BI}
({\bf p}\cdot {\boldsymbol\epsilon}^{(\alpha')})_{IA}}\over{\omega_{IA}+\omega'}}
\right)\right|^2,
\label{kh_formula}
\end{eqnarray}
where $\omega_{IA}=\omega_I-\omega_A=(E_I-E_A)/\hbar$ is the angular frequency
corresponding to the intermediate
state $I$ and the initial state $A$ (e.g. Sakurai 1967). The intermediate state
$I$ covers all the bound states $np$ and free states $n'p$, where $n'$ is the
positive real number. Here, $r_0=e^2/(m_ec^2)=2.82\times10^{-13}{\rm\ cm}$ is the classical 
electron radius with $m_e$ and $e$ being the electron mass and charge.
Note that the summation in the formula consists of a summation over infinitely many $np$ states
and an integral over continuous free states $n'p$.
 Adopting the atomic units where $\hbar=e=m_e=1$, we note that
\begin{equation}
E_n =\omega_n =-{1\over 2n^2}
\end{equation}
for a bound state $np$ and
\begin{equation}
E_n'=\omega_{n'}={1\over 2n'^2}
\end{equation} 
for a free state $n'p$.

In this work, we make no consideration on the polarization of Raman wings, 
which is deferred to a future work.
The averaging over the solid angle and polarization states yields a numerical factor of $8\pi/3$, leading to
the Thomson cross section $\sigma_{Th}=8\pi r_0^2/3$.
In the case of Rayleigh scattering, for which the initial and the final states are the
same, the cross section can be re-written as
\begin{eqnarray}
\sigma_{Ray}(\omega)
&=& \sigma_{Th} \left|
\sum_I \left({{\omega<p>_{IA}
<p>_{AI}}\over{\omega_{IA}-\omega}} 
\right.\right.
\nonumber \\ 
&-& \left.\left.  {{\omega<p>_{IA}
<p>_{AI}}\over{\omega_{IA}+\omega}}
\right)\right|^2
\end{eqnarray}
(e.g. Sakurai 1967, Bach \& Lee 2014).

The bound state radial wavefunction is written as
\begin{eqnarray}
R_{nl}(r) &=& {2\over n^{l+2}(2l+1)!} \left[
{(n+l)! \over (n-l-1)!} \right]^{1/2} (2r)^l e^{-r/n}
\nonumber \\
&&\times F(-n+l+1,2l+2, 2r/n),
\end{eqnarray}
where  $F(\alpha,\beta,z)$ is the hypergeometric function defined by
\begin{equation}
F(\alpha,\beta,z)=1+{\alpha\over\beta}{z\over 1!}+{\alpha(\alpha+1)\over
\beta(\beta+1)}{z^2\over 2!}+\cdots.
\end{equation}
For a free state $|n'l>$, the radial wavefunction is written as
\begin{eqnarray}
&&\hskip-15pt R_{n'l}(r) = {2n'^{1/2}\over [1-e^{-2\pi n'}]^{1/2}(2l+1)!}
 \prod_{s=1}^l \left(1+{s^2\over n'^2} \right)^{1/2}
\nonumber \\
&&\hskip-10pt \times (2r)^l e^{-ir/n'}F(in'+l+1, 2l+2, 2ir/n')
\end{eqnarray}
with the normalization condition
\begin{equation}
\int_0^\infty R_{n'l}(r)R_{n''l}(r) r^2 dr=\delta(n'-n'')
\end{equation}
(e.g. Bethe \& Salpeter 1957, Saslow \& Mills 1969).

With these expressions of the radial wavefunctions 
the explicit expression for relevant matrix elements between the $1s$ state
and an $np$ bound state is given by
\begin{equation}
<np|p|1s>=-i\left[{2^6 n^3 (n-1)^{2n-3}\over (n+1)^{2n+3}}
\right]^{1/2}.
\end{equation} 
Between a free $n'p$ state
and the ground $1s$ state the matrix element is given by
\begin{equation}
<n'p|p|1s>=-i\left[{2^6 n'^3 (1-e^{-2\pi n'})^{1/2}}
\over (n'^2+1)^{3}e^{2n'\tan^{-1}(1/n')}
\right]^{1/2}.
\end{equation}

In the case of Raman scattering where $A\neq B$, the Kronecker delta term
vanishes in Eq.(\ref{kh_formula})
leaving the two summation terms which contribute to the cross section.
The relevant matrix elements are
$<np|p|2s>, <np|p|3s>$ and $<np|p|3d>$ and their free state 
counterparts, which are listed in Table~1.

\begin{table*}
 \centering
  \caption{Matrix elements of the momentum operator $p$ in atomic units. }
  \begin{tabular}{@{}rl@{}}
  \hline
   Matrix & Element          \\
 \hline
$<2s|p|n p>\ =$   & $-i32\sqrt{2} n^{3/2} (n^2-1)^{1/2} (n^2-4)^{-2}
[(n-2)/(n+2)]^n
$        \\
$<2s|p|n'p>\ =$   &  $-i32\sqrt{2} n'^{3/2} (n'^2+1)^{1/2} (n'^2+4)^{-2}
(1-e^{-2\pi n'})^{-1/2} e^{-2n'\tan^{-1}(2/n')}  $   \\

 $<3s|p|n p>\ =$  &  $-i24\sqrt{3} n^{3/2} (n^2-1)^{1/2} (17n^2 - 27) 
 (n^2-9)^{-3} [(n-3)/(n+3)]^{n} $     \\
 $<3s|p|n' p>\ =$ &  $-i24\sqrt{3} n'^{3/2} (1+n'^2)^{1/2}(7n'^2+27) 
(1-e^{-2\pi n'})^{-1/2}(n'^2+9)^{-3} e^{-2n'\tan^{-1}(3/n')}
$         \\
 $<3d|p|n p>\ =$  &  $-i96\sqrt{3} n^{11/2} 
(n^2-1)^{1/2} (n^2-9)^{-3}[(n-3)/(n+3)]^n $       \\
 $<3d|p|n' p>\ =$ &  $-i96\sqrt{3} n'^{11/2} (1+n'^2)^{1/2} 
(1-e^{-2\pi n'})^{-1/2}(n'^2+9)^{-3} e^{-2n'\tan^{-1}(3/n')} $       \\
\hline
\end{tabular}
\end{table*}

The total scattering cross section $\sigma_{\rm tot}$ is given by the sum of the Rayleigh
and Raman scattering cross sections. The number of Raman scattering branching 
channels differs depending on the frequency of the incident photon. For example,
the final states of \ion{H}{1} for incident photons blueward of Ly$\beta$ may 
include $1s$, $2s$, $3s$ and $3d$ states, whereas only $1s$ and $2s$ states can be the final
state for those redward of Ly$\beta$ and blueward of Ly$\alpha$.

\begin{figure}
\includegraphics[scale=0.65]{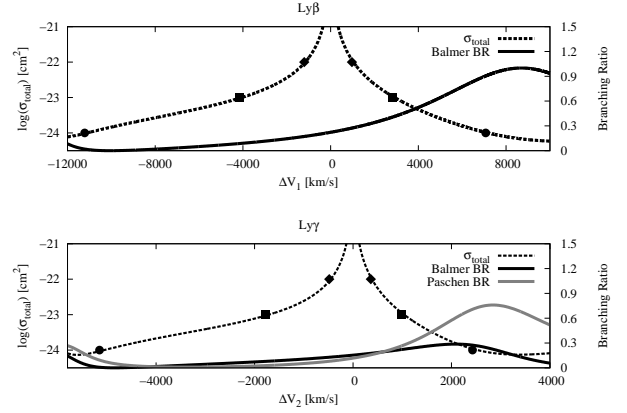}
\caption{Total scattering cross section and branching ratios around Ly$\beta$ (upper panel)
and Ly$\gamma$ (lower panel). The solid lines are the sum of cross sections of Rayleigh  
scattering and Raman scattering. The dotted lines are the branching ratio of scattering into
the $2s$ state and the thin gray line shows the branching ratio into the $n=3$ levels. Note that 
for $\lambda\ge 974.48{\rm\ \AA}$ the branching ratio into $n=3$ is larger than that into $n=2$.
}
\label{xsection}
\end{figure}

In Fig.~\ref{xsection}, we show the total scattering cross section $\sigma_{\rm tot}$ 
by a dotted line. We also show the branching
ratio into $2s$  by a thick solid line and the branching ratio into $n=3$ states 
yielding Pa$\alpha$ wings by a thin gray line.
We define the Doppler velocity factor $\Delta V_1$ around Ly$\beta$ by
\begin{equation}
{{\Delta V_1}} = {{\lambda-\lambda_\beta}\over \lambda_\beta}c
\label{delv1}
\end{equation}
where $\lambda_\beta$ is the wavelength of Ly$\beta$.
In terms of $\Delta V_1$ we find
the total scattering cross section exceeding $10^{-22}{\rm\ cm^{2}}$ in the 
range $-1,264{\rm\ km\ s^{-1}} <\Delta V_1 < +930{\rm\ km\ s^{-1}}$.
As \cite{lee2013} pointed out the total cross section is stronger in the blue
side than the red side around Ly$\beta$. 
The range of $\Delta V_1$ corresponding to
$\sigma_{\rm tot}\ge 10^{-23}{\rm\ cm^{2}}$
is $-4,162 {\rm\ km\ s^{-1}}<\Delta V_1 < +2,805{\rm\ km\ s^{-1}}$.

In a similar way, we introduce the Doppler factor $\Delta V_2$ defined as
\begin{equation}
{{\Delta V_2}} = {{\lambda-\lambda_\gamma}\over \lambda_\gamma}c
\label{delv2}
\end{equation}
where $\lambda_\gamma$ is the wavelength of Ly$\gamma$.
For radiation in the vicinity of Ly$\gamma$ the velocity
range for $\sigma_{\rm tot}\ge 10^{-22}{\rm\ cm^{2}}$ is given by
$-553{\rm\ km\ s^{-1}} <\Delta V_2 < 281{\rm\ km\ s^{-1}}$. 
 When we increase the column density
to $N_{\rm HI}=10^{23}{\rm\ cm^{-2}}$, the velocity range corresponding to the total optical 
depths exceeding unity is
$-1,809{\rm\ km\ s^{-1}}<\Delta V_2 < +934{\rm\ km\ s^{-1}}$. 
In the vicinity of Ly$\gamma$, the total cross sections are also asymmetric showing larger values in the blue part than
in the red part. This trend is similar to that around Ly$\beta$. Furthermore, the branching ratio 
of scattering into the $2s$ state is also increasing as a function of the wavelength. In particular,
the branching ratio into $n=3$ in Fig.~\ref{xsection} is larger than that into $n=2$ for $\lambda \ge 974.48{\rm\ \AA}$.
This implies that in the red part of Ly$\gamma$ with incident wavelength $\lambda \ge 974.48{\rm\
 \AA}$  we expect more Raman scattered photons redward of Pa$\alpha$
than those redward of H$\beta$. 

In Table~2 we summarize these values for neutral column densities ranging 
from $10^{22}{\rm\ cm^{-2}}$ to
$10^{24}{\rm\ cm^{-2}}$. As is shown in the table, the velocity widths
are significantly larger around Ly$\beta$ than around Ly$\gamma$. The Raman
wing width is obtained roughly from this width in the parent wavelength space
multiplied by the numerical factor $(\lambda_o/\lambda_i)^2$ due to the inelasticity
of scattering. The latter factor is the largest for Pa$\alpha$ wings and smallest
for H$\beta$ wings. From this we may expect that the extent of H$\beta$ wings
will be much smaller than that of Pa$\alpha$ wings. The branching ratio
into $n=3$ states being comparable to that into $n=2$ around Ly$\gamma$, 
the Pa$\alpha$ wings
will be much broader but shallower than the H$\beta$ counterparts.

The total scattering optical depth 
of unity for $N_{\rm HI}=10^{23}{\rm\ cm^{-2}}$ is obtained for incident radiation
with wavelengths
$\lambda_{\beta1}=1011.83{\rm\ \AA}$,
$\lambda_{\beta2}=1035.69{\rm\ \AA}$ around Ly$\beta$.
These photons are Raman scattered to appear at $6047{\rm\ \AA}$, and $7012{\rm\ \AA}$ 
around H$\alpha$.

In a similar way, we have the total scattering optical depth of unity at
$\lambda_{\gamma1}=966.99{\rm\ \AA}$,
$\lambda_{\gamma2}=975.90{\rm\ \AA}$ around Ly$\gamma$, for which Raman scattered
features reappear at $4735{\rm\ \AA}$ and $4956{\rm\ \AA}$ around H$\beta$
and at $16890{\rm\ \AA}$ and $20090{\rm\ \AA}$ around Pa$\alpha$.

\begin{table*}[t]
\centering
\begin{tabular}{l|cccc}
\hline
\hline
$N_{\rm HI}$   	   & $   \lambda_1$  & $\Delta V_1$	
& $   \lambda_2$   & $\Delta V_2$          \\  
$[{\rm cm^{-2}}]$  & $ [{\rm\ \AA}]$ & $[{\rm km\ s^{-1}}]$ 
& $ [{\rm\ \AA}]$  & $[{\rm km\ s^{-1}}]$  \\ 
\hline
$10^{22}$	& $1021.75,\ 1029.27$  & $-1264,\ +930$     & $971.07,\ 973.78$  &  $-553,\ +281$ \\   
$10^{22.5}$	& $1018.37,\ 1031.76$  & $-2253,\ +1657$    & $969.74,\ 974.62$  &  $-960,\ +539$ \\  
$10^{23}$ 	& $1011.83,\ 1035.69$  & $-4162,\ +2805$    & $966.99,\ 975.90$  &  $-1809,\ +934$ \\  
$10^{23.5}$	& $1000.30, \ 1041.51$  & $-7530,\ +4506$    & $961.69,\ 977.75$  &  $-3443,\ +1505$ \\ 
$10^{24}$ 	& $987.68, \ 1050.29$  & $-11218,\ +7069$   & $956.14,\ 980.74$  &  $-5151,\ +2427$ \\ 
\end{tabular}
\caption{Wavelengths and the corresponding Doppler factors 
having a unit total scattering optical depth
for various values of neutral hydrogen column density $N_{\rm HI}$. 
The two values of $\lambda_1$ and those of $\lambda_2$
are wavelengths around Ly$\beta$ and Ly$\gamma$, respectively.
 The Doppler factors $\Delta V_1$ and $\Delta V_2$ are defined by Eq.~(\ref{delv1})
and Eq.~(\ref{delv2}), respectively.
}
\end{table*}

\section{Monte Carlo Radiative Transfer}


We consider a neutral scattering region as a finite cylinder characterized by
the thickness and the height. The symmetry axis is chosen to be $z$ axis and
the AGN continuum source is assumed to reside at the center of the coordinate
system. 


The AGN continuum is known to be nonthermal and typically approximated by a power law  
\citep[e.g.,][]{zheng1997,van2001}. 
It  also appears that far UV continuum spectrum around Ly$\beta$ and Ly$\gamma$ is almost
flat in many AGNs. In this respect the continuum considered in this work is described by
\begin{equation}   
\lambda F_\lambda = \lambda_0 F_0
\left( 
{\lambda\over\lambda_0}
\right)^{\alpha},
\label{power_law}
\end{equation}
where the spectral index $\alpha$ is set to zero in this work,
 and $\lambda_0$ is a characteristic wavelength in this spectral
region.



Spectropolarimetry of the prototypical Seyfert 2 galaxy NGC 1068 performed by
\cite{ant1985} shows that H$\beta$ appears broad in the polarized flux.
This implies that NGC 1068 possesses the broad line region that is hidden from the
observer's line of sight by an optically thick torus-like region. 
Almost constant position angle in the polarized flux is
consistent with the scattering region located in the polar
direction with respect to the central engine. 

In performing our simulations it is assumed that the far UV incident radiation does
not affect the physical and chemical conditions of the neutral scattering region. 
With this
assumption, the Rayleigh-Raman radiative transfer of far UV continuum
can be divided coceptually into radiative transfer of each individual photon with a definite frequency
that constitutes the whole far UV continuum.

Compared with usual Raman spectroscopy performed in a lab with a monochromatic laser
with a definite polarization state, our simulation of Raman scattering by atomic hydrogen
in AGN differs mainly in our neglect of polarization. As far as no measurement is made
of the polarization of the final emergent photon, the radiative transfer using Eq.~(7)
for unpolarized light is equivalent to obtaining the total number flux
of Raman scattered radiation from simulations that take a full consideration of polarization.
With these limitations noted,
we simulate the Raman wing formation by injecting an unpolarized individual photon with a definite
frequency.
The number of these photons is determined in an accordance of AGN comtinuum spectrum.

Additional information regarding the existence of the optically thick component can be obtained from studies of X-ray 
hardness of AGNs. Because soft X-rays suffer more severe extinction than hard X-rays,
type 2 AGNs tend to exhibit larger X-ray hardness than type 1 AGNs. X-ray studies show
that the optically thick component is characterized by a hydrogen column density
ranging $10^{22-24}{\rm\ cm^{-2}}$.

\begin{figure}
\includegraphics[scale=0.27]{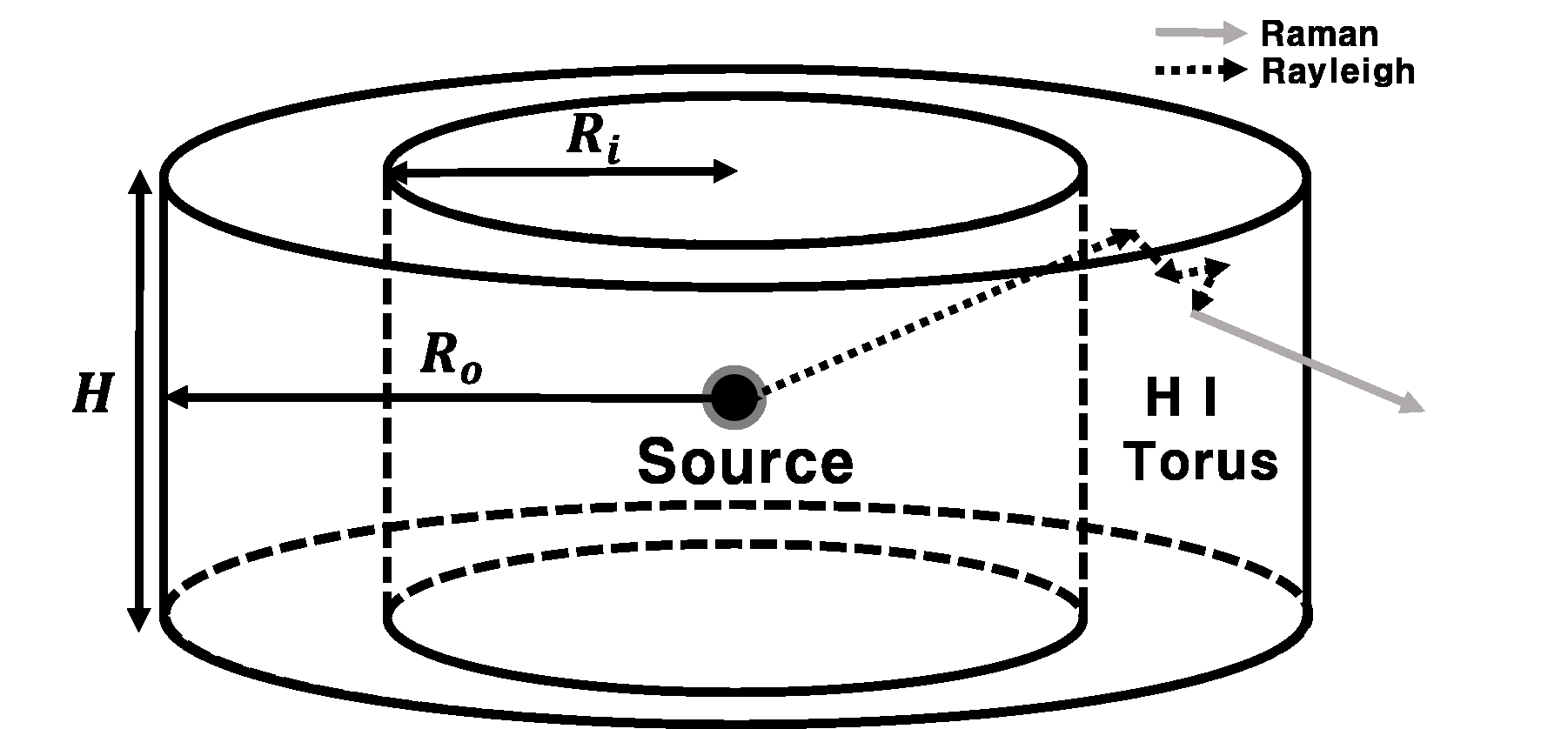}
\caption{Scattering geometry of AGNs, where the neutral scattering region is considered
to be a cylindrical torus with $R_i$, $R_o$ and $H$ being the inner radius, outer radius
and the height.
}
\label{scat_geo}
\end{figure}

In this work we place an optically thick component in the form of a cylindrical torus
with a finite height, which is schematically shown in Fig.~\ref{scat_geo}. 
This torus is specified by the inner radius $R_i$, the outer radius
$R_o$ and the height $H$. 
For the sake of simplicity, we fix the thickness of the torus $\Delta R=R_o-R_i=10\,{\rm pc}$ and we
assume that the number density of neutral hydrogen
$n_{\rm HI}$ is uniform in the scattering region. 
In this case, the scattering region can be specified by the lateral
column density $N_{\rm HI}  = n_{\rm HI}\Delta R$. The height $H$ 
is parameterized by $A=H/\Delta R$, the ratio of the height and the thickness $\Delta R$. 
In this work, we vary $A$ between 0.5 and 2 and also consider $N_{\rm HI}$ in the range  
$10^{22-24}{\rm\ cm^{-2}}$.

The Monte Carlo simulation starts with the generation of a far UV continuum photon 
from the central
engine, which subsequently enters the scattering region. The wavelength of the incident photon
is determined in accordance with the power law of the AGN continuum given in Eq.~(\ref{power_law}).
For this wavelength we rescale the scattering geometry in terms of the total scattering optical
depths $\tau_R = \sigma(\lambda)n_{\rm HI}\Delta R$ and $\tau_H=\sigma(\lambda)n_{\rm HI}H$.

To determine the first scattering site we compute the optical depth $\tau$ for this photon given by
\begin{equation}
\label{tau}
\tau=-\ln R,
\end{equation}
where $R$ is a random number uniformly distributed in the interval [0,1].

According to the branching ratio we determine the scattering type. 
If the scattering is Rayleigh,
then we look for the next scattering site by taking another step given by Eq.~(\ref{tau}). 
If the branching is into $n=2$ or $n=3$ states, then we have a Raman scattered photon which is supposed 
to escape from the region immediately.
Both Rayleigh scattering and Raman scattering share the scattering phase function
which is also identical with that of the Thomson
scattering and therefore we choose the direction
of the scattered photon in accordance with the Thomson 
phase function \citep[e.g.,][]{yoo2002,schmid89,nuss1989}.

The collection of Raman scattered photons emergent from the neutral region is made by
taking into consideration the difference between the observed wavelength space
and the parent far UV wavelength space. 
The Raman wavelength interval $\Delta\lambda_o$ corresponding to a fixed wavelength 
interval $\Delta\lambda_i$ 
of the incident radiation varies with 
$\lambda_i$ in accordance with Eq.~(\ref{wavelength_rel1}). In terms of $\lambda_i$
this relation can be recast in the form 
\begin{equation}
w_{\lambda_o}(\lambda_i) = {\Delta\lambda_i \over \Delta\lambda_o}
=\left({1- {\lambda_i\over\lambda_\alpha}}\right)^{2}.
\label{wavelengthspace}
\end{equation}
This is the factor that relates the number flux
per unit wavelength in the parent wavelength space to that in the observed (Raman scattered)
wavelength space,
which is multiplied to the Monte Carlo simulated flux for proper normalization.

\section{Monte Carlo Simulated Raman Wings}

\subsection{Balmer and Paschen Wings Formed through Raman Scattering}

\begin{figure}
\includegraphics[scale=0.65]{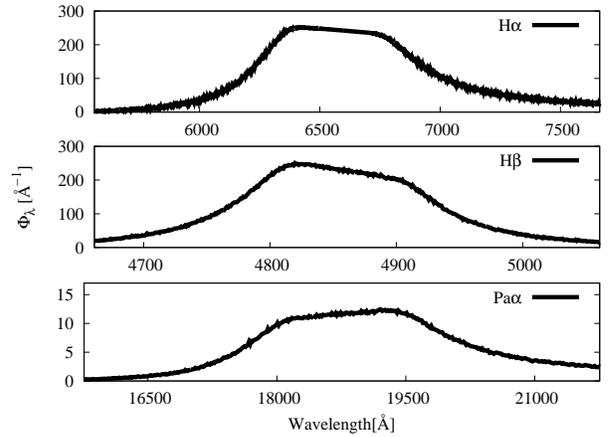}
\caption{Monte Carlo simulated profiles of Raman scattered radiation. 
The upper panel show the Raman scattering wings around H$\alpha$, H$\beta$ and Pa$\alpha$ (upper, middle and bottom panel, respectively).
The neutral scattering region is assumed to be a slab with finite thickness and infinite lateral dimensions.
The far UV source is located in the midplane.
}
\label{wings1}
\end{figure}

In Fig.~\ref{wings1}, we show the Monte Carlo simulated wings 
around H$\alpha$, H$\beta$ and Pa$\alpha$ 
formed through Raman scattering 
of far UV radiation from the central engine. 
The parameters associated with the scattering region are
$\Delta R = 10\ {\rm pc}$ and $H=\infty$, so that the far UV source is immersed at the center of a cylinder 
with an infinite height. This choice of $H=\infty$ allows us to study the basic properties of Raman
wings without complicating effects due to finite covering factors.

We set the neutral hydrogen column density $N_{HI}=n_{HI}\Delta R=10^{23}{\rm\ cm^{-2}}$ in this figure.
In this simulation, $10^4$ photons are generated per parent wavelength interval of $1{\rm\ \AA}$.
 The horizontal axis represents the observed wavelength. The vertical axis represents the number of photons obtained in the simulation per unit observed wavelength interval $\Delta \lambda_o$.

The profiles of Raman scattered radiation are characterized by the inclined central 
core part and extended wing part
that declines to zero. The inclined central part is formed from far UV photons with the total
scattering optical depth exceeding $\sim 10$. 
Roughly speaking, all the far UV photons within these ranges undergo multiple Rayleigh 
scattering processes before being converted to optical photons around the Balmer emission 
centers or IR photons around Pa$\alpha$.  
The Raman number flux  near H$\alpha$ core decreases redward 
due to the wavelength space factor given in Eq.~(\ref{wavelengthspace}), which is a decreasing
function of $\lambda_i$. 
The Raman flux around H$\beta$ shows more steeply inclined core part than that around H$\alpha$.
This is explained by the fact that as $\lambda_i$ increases the number of photons
channeled into the Pa$\alpha$ branch increases very steeply reducing the H$\beta$ flux.

However, far UV photons outside these
ranges will be scattered at most once to escape from the scattering region either as far UV photons or 
as Raman scattered photons. For incident far UV radiation 
with small scattering optical depths
the resultant wing profile will be approximately proportional to the product of the total optical depth and the branching ratio, which is in turn approximately given by the Lorentzian. 
As \cite{lee2013} pointed out,
the cross section and the branching ratios are complicated functions of wavelength, for
which a quantitative investigation can be effectively performed
adapting a Monte Carlo technique.

\begin{figure}
\includegraphics[scale=0.65]{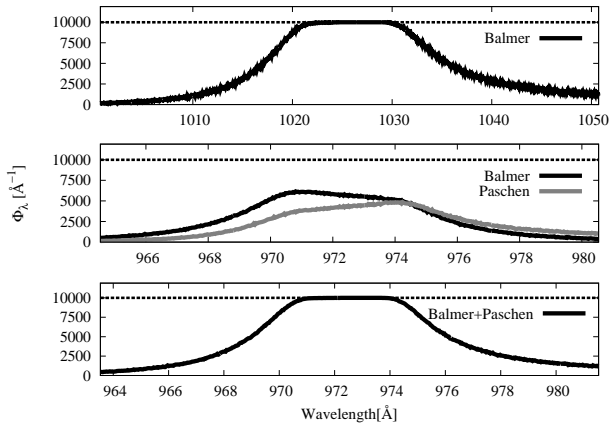}
\caption{Monte Carlo simulated profiles of Raman scattered radiation in the parent
wavelength space. The horizontal dotted line shows the incident flux taken to be $10^4$ photons per unit wavelength interval.
The top panel shows the Raman scattered radiation around Ly$\beta$
which shows flat core part because of the complete Raman conversion of
the flat incident radiation.
The mid panel shows the Raman scattered profiles that appear near H$\beta$ (black lines)
and near Pa$\alpha$ (gray lines) from the flat UV radiation near Ly$\gamma$. 
The bottom panel shows the sum of Raman scattered fluxes around H$\beta$ and Pa$\alpha$
transformed into the parent wavelength space .
}
\label{wings2}
\end{figure}

In Fig.~\ref{wings2}, we transform the same Monte Carlo data given in Fig.~\ref{wings1}
to the parent
wavelength space in order to check easily the fraction of Raman scattered photons
with respect to the incident radiation. 
The horizontal dotted line show the incident flux taken to be $10^4$ photons per unit wavelength interval.
The top panel shows the number of Raman
scattered radiation around H$\alpha$. The core part is flat and coincides with the incident flux, which verifies that the
Raman conversion is almost complete. 

The mid panel shows Raman scattered H$\beta$
and Pa$\alpha$ wings by a black solid line and a gray solid line, respectively. 
The bottom panel shows the sum of Raman scattered H$\beta$
and Pa$\alpha$, where we recover the flat core part coinciding with the incident radiation.
This fact provides confirmation that the Raman conversion is also complete
in the vicinity of the Ly$\gamma$ core.
In this figure, it can be seen that more Pa$\alpha$ wing photons are obtained 
  than H$\beta$ wing photons for incident wavelength $\lambda>974.48{\rm\
\AA}$, where the branching ratio into $n=3$ states exceeds that into the $2s$ state.

\subsection{Asymmetry of Raman Scattering Wings}

\begin{table}
\centering
\begin{tabular}{l|ccc}
\hline
$N_{\rm HI}\,$   	   & $ \eta_{\rm H\alpha} $  & $ \eta_{\rm H\beta}$	& $\eta_{\rm Pa\alpha}$  \\
$ \rm cm^{-2}$  \\   
\hline
$10^{22}$	& $0.86$ & $1.13$ & $0.86$ \\	
$10^{22.5}$	& $0.81$ & $1.21$ & $0.79$ \\	
$10^{23}$ 	& $0.73$ & $1.35$ & $0.69$ \\	
$10^{23.5}$ 	& $0.59$ & $1.57$ & $0.55$ \\	
$10^{24}$ 	& $0.36$ & $1.76$ & $0.37$ \\	
\end{tabular}
\caption{
The ratio $\eta$ of the Raman photon number flux blueward of line center to that redward
of line center in Fig.~\ref{wings1}. 
}
\end{table}

In this subsection, we quantify the asymmetry in the wing profiles formed 
through Raman scattering around H$\alpha$, H$\beta$ and Pa$\alpha$. One way 
to do this is to compute the ratio $\eta$ of the number of Raman 
photons blueward of line center to that redward of line center.
In Table~3 we show our result.

In the case of H$\alpha$, the red wing part extends further away from the H$\alpha$ center 
than the blue wing, which is clearly seen by the fact that $\eta_{\rm H\alpha}$ is less
than 1. This phenomenon is mainly due to the higher branching ratio 
for photons redward of Ly$\beta$ than their blue counterparts. A similar behavior is also
observed for the Pa$\alpha$ wings due to increasing branching ratio redward of Ly$\gamma$.

However, in the case of H$\beta$ we note that $\eta_{\rm H\beta}$ exceeds unity
implying that the blue wing is stronger than the red part. 
This behavior is attributed to the much stronger
total cross section in the blue part than in the red part near Ly$\gamma$, despite the slow increase of the branching ratio as wavelength.

Another way of investigating asymmetry is to identify the half-value locations
from a reference Raman flux value.
Taking the simulated Raman flux per unit wavelength 
at the H$\alpha$ line center as a reference value, 
the half values appear at $-11,900{\rm\ km\ s^{-1}}$ and $+12,800{\rm\ km\ s^{-1}}$.
A similar analysis for H$\beta$ wings shows that the corresponding Doppler factors
are  $-4,330{\rm\ km\ s^{-1}}$ and $+3,500{\rm\ km\ s^{-1}}$. Pa$\alpha$ wings
are much wider and the Doppler factors corresponding to the half-core values 
are  $-13,500{\rm\ km\ s^{-1}}$ and $+17,000{\rm\ km\ s^{-1}}$.

\subsection{Dependence on the Scattering Geometry}

\begin{figure}
\includegraphics[scale=0.65]{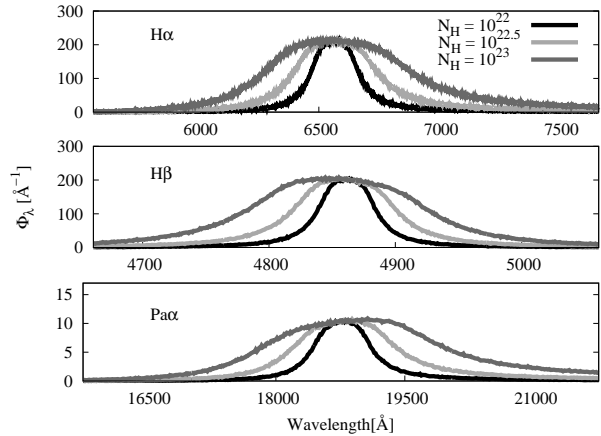}
\caption{Monte Carlo simulated profiles of Raman scattered radiation
around H$\alpha$, H$\beta$ and Pa$\alpha$ (upper, middle and bottom panel, respectively) 
for various column densities $N_{HI}$ ranging from $10^{22}{\rm\ cm^{-2}}$ to $10^{23}{\rm\ cm^{-2}}$.
As $N_{HI}$ increases, the wing profile broadens in roughly proportional to $N_{HI}^{1/2}$.
}
\label{wings_col}
\end{figure}

In Fig.~\ref{wings_col}, we investigate the Balmer and Paschen wings
that are simulated for various 
values of $N_{\rm HI}$.  In this figure, we set $R_i=10{\rm\ pc}$ 
and $H=R_o=20{\rm\ pc}$.  We vary the number density of H~I in such a way that
the neutral column density $N_{\rm HI}$ in the lateral direction becomes
$10^{22},\ 10^{22.5}$ and $10^{23}{\rm\ cm^{-2}}$. 

Because the covering factor of the neutral scattering region is fixed and
the total scattering optical depth is very large near line core, the Raman flux
at line center remains the same for various values of $N_{\rm HI}$.
The primary effect of varying $N_{\rm HI}$ is the width of the Raman wings, in that
the width is roughly proportional to $N_{\rm HI}^{1/2}$. As $N_{\rm HI}$ increases,
the saturated part also extends further away from the line center.

\begin{figure}
\includegraphics[scale=0.65]{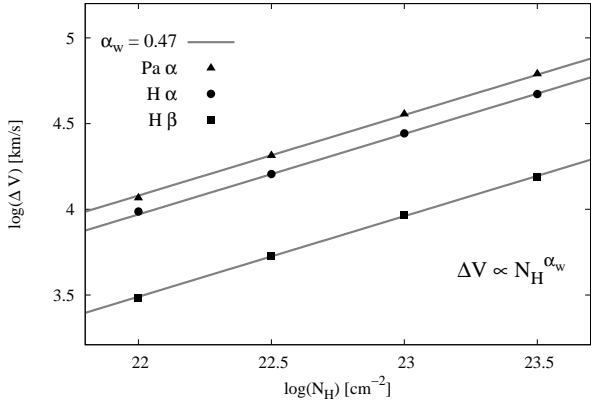}
\caption{Widths of Monte Carlo simulated Raman wings around H$\alpha$, 
H$\beta$ and Pa$\alpha$ considered in Fig.~\ref{wings_col}. The horizontal and vertical 
axes represent the column density and the wing width, respectively. Both axes are 
in logarithm scale. The dots are measured values from the Monte Carlo
simulation and the fitting lines have a slope of 0.47.
}
\label{wings_width}
\end{figure}

We define the width of Raman wings
as the difference of the two Doppler factors that correspond to the half
values of the Raman flux at line center. Using this definition, we plot
the Raman wing widths for various column densities in Fig.~\ref{wings_width}.
Note that both the vertical and the horizontal scales are logarithmic.
The Monte Carlo
simulated data are shown by the dots that are fitted by lines having a slope of 0.47.
In terms of the parameter $N_{23}=N_{\rm HI}/(10^{23}{\rm\ cm^{-2}})$
the fitting lines are explicitly written as
\begin{eqnarray}
\Delta V_{\rm H\alpha} &=& 2.75\times10^4\ N_{23}^{0.47}\ {\rm km\ s^{-1}} 
\nonumber \\
\Delta V_{\rm H\beta} &=& 9.12\times 10^3\ N_{23}^{0.47} \ {\rm km\ s^{-1}}
\nonumber \\
\Delta V_{\rm Pa\alpha} &=& 3.55\times 10^4\ N_{23}^{0.47}\ {\rm km\ s^{-1}}.
\end{eqnarray}

%
%
%

\begin{figure}
\includegraphics[scale=0.65]{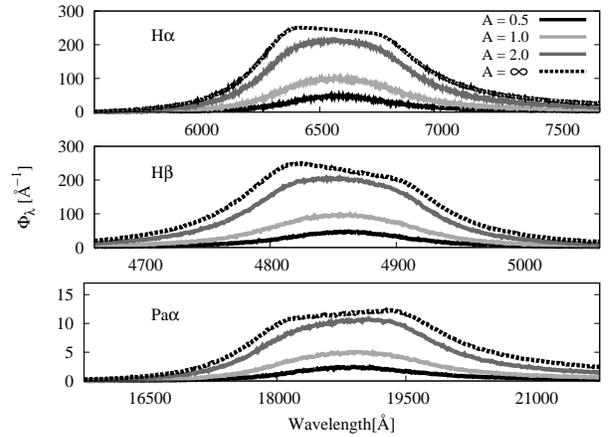}
\caption{Monte Carlo simulated profiles of Raman scattered radiation 
around H$\alpha$, H$\beta$ and Pa$\alpha$ (upper, middle and bottom panel, respectively) 
for various covering factors.  $A=H/\Delta R$ is the ratio of the height
and the thickness of the cylindrical shell.
}
\label{wings_cov}
\end{figure}

In order to investigate the effect of the covering factor of the scattering region, 
we vary the height $H$ of the scattering region with the parameters 
$R_i=10{\rm\ pc}$, $R_o=20{\rm\ pc}$ and $N_{\rm HI}=10^{23}{\rm\ cm^{-2}}$ fixed. 
We show our result in Fig.~\ref{wings_cov} for values of $A=0.5, 1, 2$ and $\infty$.

\begin{figure}
\includegraphics[scale=0.65]{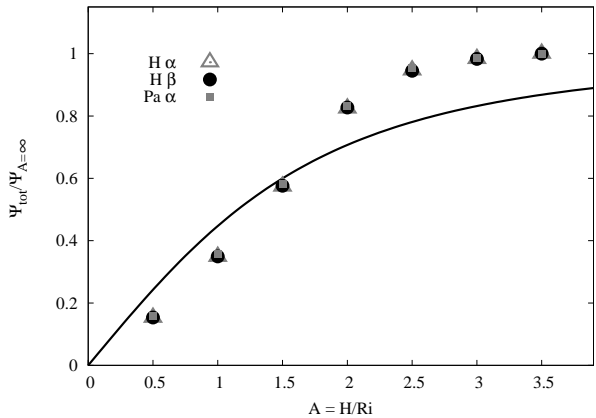}
\caption{Relative strengths of Monte Carlo simulated Raman wings around H$\alpha$, 
H$\beta$ and Pa$\alpha$ considered in Fig.~\ref{wings_cov}. The vertical axis
represents the wing strengths normalized with the Raman wing strength for $A=\infty$.
The solid curve shows the covering factor $f(A)={A\over\sqrt{4+A^2}}$ of the
scattering region, whose behavior is also characterized by linearity for small $A$ and 
approaching unity as $A\to\infty$.
}
\label{width_cov}
\end{figure}

With $A\le 2$ the wing profiles are smooth. Even though not shown in the figure,
we checked that there appears an inclined plateau around line center in each
Raman wing for $A\ge 4$, implying the saturation behavior.

As the covering factor increases, more far UV photons are incident
into the scattering region leading to stronger Raman scattering wings.
When $A$ is small, the Raman wing strength is approximately linearly proportional to $A$.
However if $A$ exceeds 2, the wing strength increases very slowly and 
approaches a limiting value. 
Given $A$ we normalize the H$\alpha$ Raman wing strengths 
by dividing the total number of Raman H$\alpha$ wing photons by that for $A=\infty$.
The wing strengths for H$\beta$ and Pa$\alpha$ are normalized
in a similar way.
In Fig.~\ref{width_cov}, we plot the normalized wing strengths for
H$\alpha$, H$\beta$ and Pa$\alpha$ by triangles, circles and squares, 
respectively. Given $A$ and $N_{HI}$, the normalized strengths
are almost equal to each other for H$\alpha$, H$\beta$ and Pa$\alpha$ wings.
For reference,  we add the solid curve to show the covering factor 
of the scattering region given by
\begin{equation}
f(A) = {A\over\sqrt{4+A^2}}.
\end{equation}
The covering factor also
shows a similar behavior of  linearity for small $A$ and approaching
unity for large $A$. Significant deviations between the normalized Raman wing
strengths and the covering factor are
attributed to complicated
effects of multiple scattering and branching channels associated with the
formation of Raman wings.

%
%
%
%

\subsection{Mock Spectrum around Balmer Lines}

\begin{figure}
\includegraphics[scale=0.65]{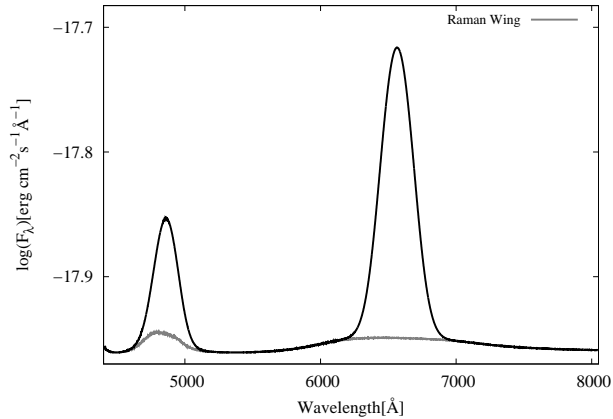}
\caption{Mock composite spectrum around H$\alpha$ and H$\beta$ with Raman
scattering wings. 
The gray line shows Raman wings generated through a Monte Carlo simulation. The vertical scale is logarithmic for clear view of Raman wings against
prominent emission lines.
}
\label{mock_spect}
\end{figure}

In order to assess the observational feasibility we produce a mock
spectrum around Balmer lines by superposing the wing profiles onto
artificially produced H$\alpha$ and H$\beta$ broad emission lines.
In Fig.~\ref{mock_spect}, we show our result.
In the production of the mock spectrum 
we assumed that the continuum level around Ly$\beta$ and Ly$\gamma$
is given by the fixed value of $\lambda L_\lambda
=10^{44}{\rm\ erg\ s^{-1}}$ and that the continuum between
H$\beta$ and H$\alpha$ is given in such a way that
$L_\lambda = 10^{40}{\rm\ erg \ s^{-1} \AA^{-1} }$.
We also set the luminosity distance $D_L=0.3{\rm\ Gpc}$.

With no established line profiles of the Balmer emission lines in AGNs, 
we take a Gaussian profile of width $5,000{\rm\ km\ s^{-1}}$ 
for H$\alpha$ and H$\beta$ emission lines. We set the equivalent widths
of H$\alpha$ and H$\beta$ broad emission lines to be 200 \AA\
and 50 \AA, respectively.
The scattering geometry is taken so that 
$R_i=10\:{\rm pc}, R_o=20\:{\rm pc}$ and $H=20\:{\rm pc}$ with
$N_{\rm HI}=5 \times 10^{23}\:{\rm cm^{-2}}$. 

In the figure the vertical scale is logarithmic and the Raman wings
are shown by gray solid lines. 
In the case of H$\beta$, the Raman
wings are inconspicuous because the broad emission component dominates
the relatively narrow Raman H$\beta$ wings. However, the Raman
H$\alpha$ wings are sufficiently wide to be observationally detectable.
It is an interesting possibility that type 2 AGNs may also show 
detectable Raman H$\beta$ wings, which will be seen outside of the
narrow H$\beta$ emission line.

\section{Summary and Discussion\label{sec:format}}

In this article we produced Raman wing profiles expected around H$\alpha$, H$\beta$
and Pa$\alpha$ that are formed by far UV continuum radiation scattered in a thick
neutral region surrounding the central engine. The strengths of
Raman wings are mainly determined by the product of the covering factor
and the neutral column density of the scattering region. The wing width
is approximately proportional to $N_{\rm HI}^{1/2}$. 
We also provide a mock spectrum 
by superposing simulated Raman wings onto artificially generated broad
emission lines of H$\alpha$ and H$\beta$. 

 Observationally the Raman wings are difficult to discern from the underlying
continuum because of their large width. Another issue may be that broad wings
can also be formed from Thomson scattering or hot tenuous fast wind 
that emits Balmer and Paschen line
photons (\cite{kim07}). One distinguishing aspect of the Raman
wings is the differing widths and profiles exhibited by H$\alpha$, H$\beta$ and Pa$\alpha$
because of the complicated atomic physics. If the wings are formed in an emission
region moving very fast, then all the wings are expected to show similar widths and profiles.

Another important characteristic is the linear polarization because the scattering mechanism is exactly
the E-1 process, which also characterizes the Thomson scattering (\cite{trippe14}).
\cite{ogle97} performed spectropolarimetry of the prototypical narrow line radio galaxy Cyg~A 
using the Keck~II Telescope, in which
they discovered extremely broad H$\alpha$ in the polarized flux. The full width at half-maximum 
of polarized H$\alpha$ is $26,000{\rm\ km\ s^{-1}}$. If  this huge width is attributed
to dust scattering or free electron scattering,
one may need to assume that the velocity scale of the hidden broad line region is roughly the same value of $26,000{\rm\ km\ s^{-1}}$. In this case, the relativistic beaming inevitably 
leads to very asymmetric profiles with the blue part much stronger than
the red part.  

On the other hand, very broad features around H$\alpha$ are naturally formed 
through Raman scattering of far UV radiation without invoking
extreme kinematics in the broad emission line region. For example, if we assume that
the central engine of Cyg~A is surrounded by a cold thick region with $N_{\rm HI}\sim
10^{23}{\rm\ cm^{-2}}$ the width of $26,000{\rm\ km\ s^{-1}}$ can be
explained. In this case, we have to assume that the continuum around Ly$\gamma$ is
relatively weak compared to that around Ly$\beta$ in order to explain no detection
of the polarized broad features around H$\beta$. In particular, it is highly noticeable
that \cite{rey2015} proposed a neutral column density of $\sim 1.6\times10^{23}{\rm\ cm^{-2}}$
from their observations of {\it NUSTAR} of Cyg~A.
Despite the difficulty in identifying the broad wings around \ion{H}{1} emission lines from the local
continuum, they will provide important clues to the unification model of AGNs.



\acknowledgments

The authors are very grateful to the anonymous referee, who provided helpful comments
for the improvement of the current presentation.
 This research was supported by the Korea Astronomy and Space Science Institute
under the R\&D program(Project No. 2015-1-320-18) supervised by the Ministry of Science, ICT and Future Planning.

\clearpage


\end{document}